# Spatial and temporal distribution of stratospheric turbulence from global high-resolution radiosonde data

Han-Chang Ko, Hongwei Sun* (hongwei8@hawaii.edu)

Department of Atmospheric Sciences, School of Ocean and Earth Science and Technology, University of Hawai'i at Manoa, Honolulu, HI, United States

**Abstract**

Stratospheric turbulence is difficult to observe, yet it strongly affects how momentum, trace gases, and aerosols spread through the atmosphere. Here, we use global high-resolution radiosonde observations from 370 stations during October 2014–December 2025 to estimate stratospheric turbulent diffusivity with a Richardson-number-based method. We find that turbulent diffusivity spans a broad range and is dominated by turbulence occurring in statically stable but strongly sheared conditions. Enhanced values appear over Turkey, India, Malaysia, Japan, and major mountainous regions, with evidence that mountain waves and convective activity both contribute. We also identify a local maximum just above the tropical tropopause, between the Equator and 15°N at 17 km altitude, which may indicate an injection region conducive to the rapid aerosol dispersion for stratospheric aerosol injection (SAI). Stratospheric turbulent diffusivity also shows a significant increase ($3.5 \times 10^{-3}$ m$^2$ s$^{-1}$ year$^{-1}$) over 2015–2025. Finally, these results may help constrain and predict the dispersion of stratospheric plumes from rockets, aircraft, and potential SAI.

# 1. Introduction

Atmospheric turbulence is a fundamental process in the atmosphere because it controls the mixing of trace gases and the diffusion of heat and momentum across a wide range of scales of atmospheric motions. It also has practical importance for numerical weather prediction, climate modeling, and aircraft operations. Because turbulence is intermittent and spans a wide range of cascading scales that are difficult to resolve explicitly in global models, its effects are commonly represented through parameterized quantities such as turbulent kinetic energy and turbulent diffusivity (1, 2, 3). Reliable observational constraints are therefore essential for understanding turbulence characteristics and for improving model parameterizations.

Among the various atmospheric layers, stratospheric turbulence is often the most overlooked because it is difficult to observe in the stratosphere and its intensity is relatively weak compared to that in the boundary layer. However, there has been increasing interest in stratospheric turbulence because of its role in the transport and mixing of momentum and trace/aerosol constituents in the middle atmosphere. Quantitative information on stratospheric turbulent diffusion is also increasingly relevant to recent discussion of geoengineering, especially stratospheric aerosol injection (SAI) and cirrus cloud thinning (CCT), for which the upper troposphere and lower stratosphere (UTLS) is the primary target region (4, 5, 6). When using aircraft to inject aerosols or their precursor into the stratosphere for SAI, turbulent diffusivity is one of the key parameters controlling the spreading of aircraft plumes (6, 7) and therefore affects aerosol concentration and aerosol size distribution inside the near-field aerosol plumes. Despite this importance, the role of turbulent diffusion has received relatively little direct observational attention in SAI-related discussions, largely because reliable global observations of stratospheric turbulence have been scarce. As a result, serious discussion of observationally constrained stratospheric turbulent diffusivity in the context of SAI has remained limited. More broadly, both general circulation modeling of the atmosphere (8) and quantitative assessment of SAI require observationally based reference values of turbulent diffusion in the stratosphere (7). Effective diffusivity has also been used as a diagnostic of tracer transport and mixing in the lower stratosphere. For example, Haynes and Shuckburgh (9) showed that minima in effective diffusivity identify transport barriers associated with the extratropical tropopause and subtropical jet, while Nakamura (10) highlighted the broader use of eddy/effective diffusivity as a mixing diagnostic. Although those studies focused on large-scale tracer transport, they underscore the importance of diffusivity-based perspectives in stratospheric mixing.

A variety of observing platforms have been used to investigate atmospheric turbulence, including commercial aircraft, research aircraft, radar, lidar, and radiosondes. Among these, global operational high vertical-resolution radiosonde data (HVRRD) provide a unique advantage for turbulence research. HVRRD offers direct in situ observations of temperature and horizontal wind up to approximately 30 km altitude, with 1 and 2 s sampling intervals corresponding to about 5 and 10 m vertical resolution. The dataset is recently available over near-global regions for more than a decade through operational radiosonde stations (11), without requiring additional field-campaign costs. Commercial aircraft observations have provided highly valuable in situ information on atmospheric turbulence, particularly in the upper troposphere near typical flight levels (12, 13). High-altitude aircraft observations have also offered unique insights into stratospheric turbulence and related processes, although such measurements are generally obtained during intensive field campaigns and therefore are not usually available with broad spatial and temporal continuity (14, 15, 16). Remote-sensing techniques such as radar and lidar provide important observational constraints on stratospheric structure and variability, but their coverage is often limited to specific locations and their interpretation depends on the characteristics of the retrieval method (17). For these reasons, HVRRD provides an especially suitable observational basis for global analysis of stratospheric turbulence (18, 19).

Previous studies have demonstrated the usefulness of HVRRD for turbulence research in the free atmosphere, including the stratosphere. For example, Clayson and Kantha (20) suggested that HVRRD can be used to infer turbulence and mixing in the free atmosphere, while Ko *et al.* (21) further applied HVRRD over the United States mainland to characterize turbulence structures in both the troposphere and the stratosphere. Ko and Chun (22) used operational HVRRD in the United States to estimate the turbulent energy dissipation rate ($\varepsilon$) and investigated potential sources of turbulence using reanalysis-based diagnostics. More recently, Ko et al. (18) presented the first quasi-global distribution of free-atmospheric turbulence from operational HVRRD, including the z = 18–30 km, and Ko and Chun (19) proposed a Richadson number-based method that extends turbulence detection to both overturning and statically stable but shear-unstable conditions, and provided global distributions of $\varepsilon$ both in the troposphere and in the stratosphere. However, most of these previous HVRRD studies focused primarily on turbulence occurrence, layer thickness, and $\varepsilon$, whereas turbulent diffusivity has received much less direct attention, especially its comprehensive spatial and temporal variations in the stratosphere.

Previous studies have also provided important insight into stratospheric turbulence from aircraft observations. Schumann *et al.* (23) estimated plume-diffusion parameters for aircraft exhaust near the tropopause and showed that vertical mixing is very weak under stably stratified conditions, while Dürbeck and Gerz (24) used large-eddy simulations to examine aircraft-plume diffusion in weak, decaying turbulence and obtained similarly small effective vertical diffusivities at later stages of plume evolution. More recently, Dörnbrack *et al.* (15) found from high-resolution research-aircraft observations that turbulence in the upper troposphere and lower stratosphere was generally rare and mostly linked to enhanced vertical shear rather than to overturning gravity waves. Likewise, Woiwode *et al.* (16) documented non-orographic gravity waves and clear-air turbulence generated by merging jet streams and showed that observed turbulence episodes coincided with regions of low Richardson number. These studies collectively demonstrate that stratospheric turbulence is closely tied to strong shear, gravity-wave activity, and jet-related dynamics, but they were mainly based on aircraft observations, plume analyses, or individual events and therefore did not provide a global observational description of stratospheric turbulent diffusivity.

Motivated by these considerations, this study investigates the characteristics of turbulent diffusivity in the stratosphere using global HVRRD. Specifically, we analyze all currently available data, spanning 11 years and 3 months of data from October 2014 to December 2025, to document the global distribution of stratospheric turbulent diffusivity, its dependence on turbulence regime, its geographical and seasonal variability, its latitude–altitude structure, and its recent temporal evolution. We also discuss the potential implications of the observed turbulence structure for SAI's aerosol dispersion. By providing observationally based reference information on stratospheric turbulent diffusivity, this study aims to contribute to future turbulence modeling/parameterization in Lagrangian plume or box model, large-scale atmospheric applications, and discussion of turbulent diffusion in SAI-related research.

## 2. Results

## 2.1. Global occurrence distributions of stratospheric turbulence

To characterize the global distribution of turbulent diffusivity $K$ in the stratosphere ($z$ = tropopause–30 km), Fig. 1 presents the occurrence-number distributions of $K$ together with those of the variables appearing in its estimation formula using the first-order Smagorinsky closure: $K = (0.2L)^2 |Def| \sqrt{0.25 - \overline{Ri_{min}}}$ (see Eq. 1 in Materials and Methods), namely the turbulence length scale ($L$), vertical wind shear (VWS), and Richardson number (Ri). In each panel, blue and red lines denote the distributions for positive-Ri (PosRi) and negative-Ri (NegRi) cases, respectively, while the black line indicates the distribution for all cases combined. The distribution of $\log 10 K$ in Fig. 1(a) spans a range from approximately -0.5 to 2 and exhibits a positively skewed, quasi-normal shape. For all cases, the mean and median values of $\log 10 K$ are 0.554 and 0.490, respectively, which correspond to 3.58 and 3.09 m2 s-1 in linear scale. These values are consistent with previous estimates based on HVRRD and the Thorpe method. For example, Bellenger et al. (25), using Thorpe-derived diffusivity estimates over the tropical open ocean, reported $K$ values ranging from about 1 to 100 m2 s-1, with a peak near 5–10 m2 s-1. Dürbeck and Gerz (24), based on large-eddy simulations of aircraft exhaust plumes in the free stably stratified atmosphere near cruise altitudes under shearless conditions after aircraft-induced turbulence had decayed, reported vertical diffusivities ranging from 2.3 to 0.37 m2 s-1 during the initial diffusion stage and approximately 0.15 m2 s-1 at later times. Schumann et al. (23), using in situ aircraft observations of exhaust plumes encountered 5–100 min after emission at 9.4–11.3 km near the tropopause in the North Atlantic flight corridor in October 1993, estimated vertical diffusivities of 0–0.6 m2 s-1. Thus, the present estimates are close to those of Bellenger et al. (25) and to the initial values of $K$ reported by Dürbeck and Gerz (24), but are substantially larger than the later-stage values reported by Dürbeck and Gerz (24) and those reported by Schumann et al. (23). The agreement with Bellenger et al. (24) may be attributed to the fact that both studies used HVRRD and diagnosed relatively localized, burst-like, active turbulence. The consistency with the initial values of Dürbeck and Gerz (24) may reflect the fact that those values correspond to the early stage of plume evolution, when turbulence was still relatively strong. In contrast, the later-stage values reported by Dürbeck and Gerz (24) and those of Schumann et al. (23) are smaller because they primarily represent aircraft exhaust-plume diffusion under weakly turbulent conditions after the decay of aircraft-induced turbulence, rather than active turbulent layers detected from high-resolution radiosonde profiles. To further examine the vertical dependence of Fig. 1(a), we additionally calculated the occurrence-number distributions of $\log 10 K$ for successive 3 km altitude bins above the tropopause (not shown). Although the mean $\log 10 K$ increases above about 18 km from the tropopause, turbulence occurrences in that altitude range are relatively infrequent. As a result, the bulk distribution in Fig. 1(a) remains dominated by the more frequently observed lower-stratospheric turbulence, and the overall mean value does not differ substantially.

When the distributions are separated by Ri regime, marked differences emerge. Among the total 4,751,228 turbulence cases identified in the stratosphere, 3,795,786 cases (80%) belong to the PosRi regime, whereas 955,442 cases (20%) belong to the NegRi regime. This PosRi dominance is much greater than that reported for the troposphere in Ko and Chun (19), where the two regimes contributed nearly equally. Such a difference is physically reasonable because the stratosphere is more statically stable than the troposphere, which suppresses overturning turbulence and favors shear-driven turbulence. The mean $\log_{10} K$ values are 0.536 for PosRi cases and 0.626 for NegRi cases. This indicates that, although NegRi cases are less frequent, turbulence that develops into overturning against the strongly stable stratospheric background tends to be stronger than turbulence in PosRi cases.

The distributions of the variables constituting $K$ help explain the behavior of $K$. First, the occurrence distribution of $L$ in Fig. 1(b) decreases approximately exponentially with increasing value. Accordingly, this distribution is more appropriately characterized by percentile statistics than by an arithmetic mean. For all cases, the 50th, 95th, and 99th percentile values are 80, 175, and 250 m, respectively. The majority of turbulence layers with $L < 200$ m belongs to the PosRi regime, but their relative contribution decreases as L increases. Above $L \sim 350$ m, NegRi cases account for more than half of all cases. This suggests that shallower turbulence layers are dominated by developing shear-driven turbulence in the $0 < Ri < 0.25$ regime, whereas deeper turbulence layers more often correspond to mature overturning turbulence. The percentile values for PosRi cases are 80, 165, and 235 m at the 50th, 95th, and 99th percentiles, respectively, while those for NegRi cases are 85, 205, and 305 m, respectively.

Second, VWS in Fig. 1(c) exhibits an approximately normal distribution, with mean values of 0.0316, 0.0345, and 0.0203 s$^{-1}$ for all, PosRi, and NegRi cases, respectively. The relatively weak VWS in NegRi cases may indicate that these cases include overturning, convective, or buoyancy-driven turbulence in weak-shear environments; however, some NegRi cases may also reflect turbulence that has evolved from an initial shear-driven stage into overturning and subsequently into a more mature or decaying state in which mixing has already weakened the original shear. This further implies that the larger mean $\log_{10}K$ in NegRi cases is primarily attributable to their larger $L$, rather than to weaker VWS. Finally, the distribution of Ri in Fig. 1(d) is bounded above by 0.25 by the Eq. (1), and clearly shows the predominance of PosRi cases over NegRi cases in the stratosphere. Overall, these results indicate that most stratospheric turbulence occurs in statically stable layers with sufficiently strong shear, while the less frequent overturning turbulence tends to be associated with larger layer thickness and stronger diffusivity.

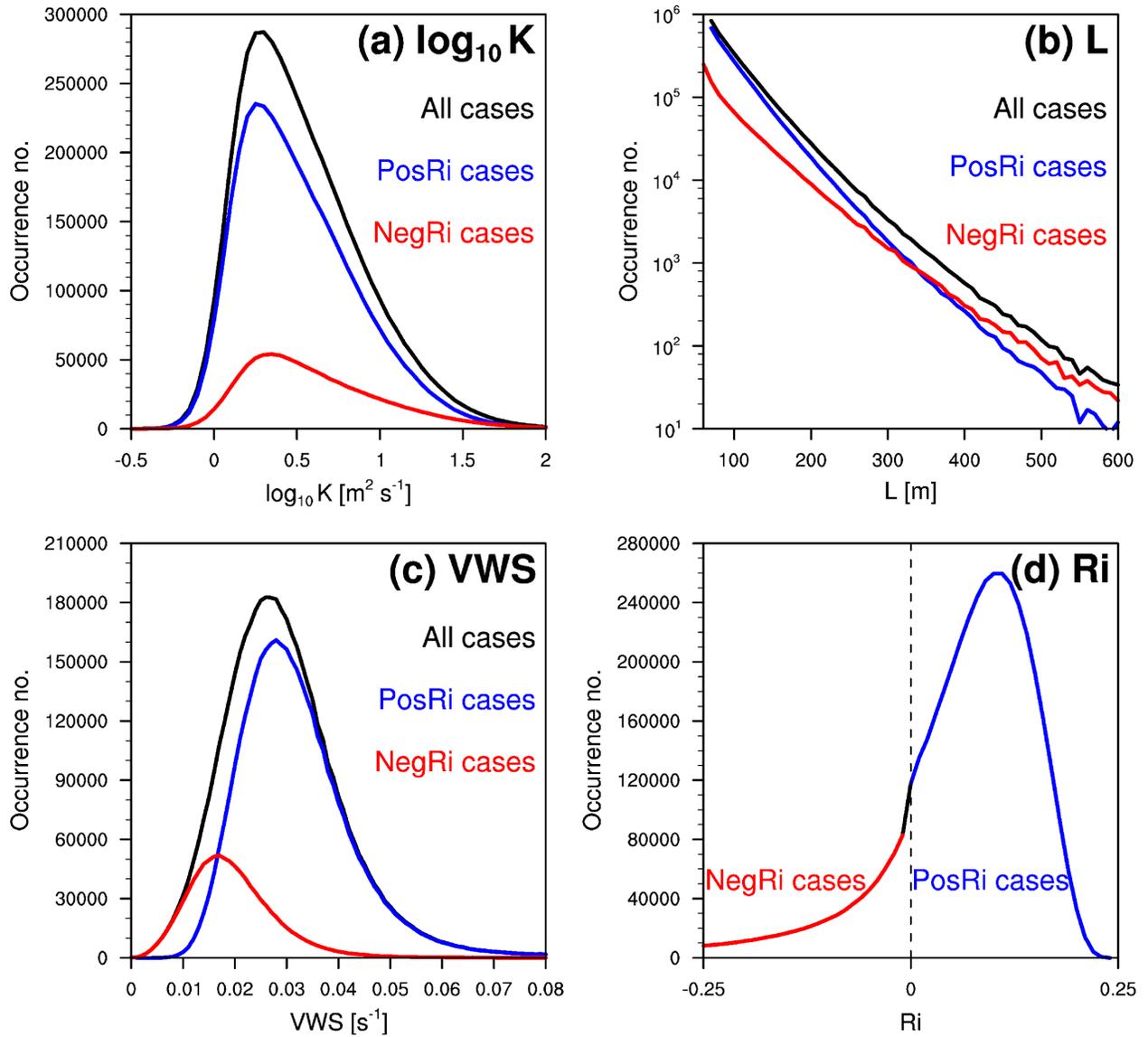

**Fig. 1. Global occurrence-number distributions of stratospheric turbulence characteristics.**
**(A)** $\log_{10} K$, **(B)** $L$, **(C)** VWS, and **(D)** Ri for z = tropopause–30 km during October 2014–December 2025. Black, blue, and red curves represent all cases, positive-Ri cases, and negative-Ri cases, respectively.

### 2.2. Horizontal distribution of stratospheric turbulence and its potential sources

Fig. 2 shows the horizontal distributions of station-mean $K$ in the stratosphere and those of the mean variables contributing to $K$. $\log_{10} K$ in Fig. 2(a) exhibits prominent maxima over Turkey, India, Malaysia, and Japan. Relatively large values are also found over major mountainous regions, including the Rocky and Appalachian Mountains in the United States and the Andes in South America. In contrast, comparatively small values are generally observed over high latitudes in both hemispheres. To examine the factors responsible for this pattern, Fig. 2(b) presents the horizontal distribution of $L$. Its overall spatial pattern is broadly similar to that of $\log_{10}K$, and most of the regions with enhanced log10$K$ also show large $L$ values. This correspondence suggests that $L$ plays a dominant role in determining the global distribution of $\log_{10}K$.

The distribution of VWS in Fig. 2(c) shows relatively large values mainly over regions of elevated topography, such as the Rocky and Appalachian Mountains, Turkey, and India. In particular, the maxima over Turkey and India, together with the enhanced $L$ values, appear to contribute to the large $\log_{10}K$ values in these regions. However, although $\log_{10}K$ is also large over Malaysia and Japan, VWS is not particularly enhanced there. Considering Fig. 2(d), Ri in these regions tends to be relatively low and negative, suggesting that the turbulence there is more strongly associated with buoyancy-driven or overturning conditions than with shear-driven turbulence. This may explain why turbulent diffusivity is relatively large despite modest VWS, implying that $L$ is the primary controlling factor in those regions. These results further indicate that the relative contributions of $L$ and VWS to $K$ vary geographically. Finally, Fig. 2(d) shows that Ri tends to be relatively low in the major regions where $\log_{10}K$ is large. This spatial correspondence is consistent with Eq. (1), in which $K$ is inversely related to Ri, and indicates that stronger instability (i.e., smaller Ri) is associated with stronger turbulence.

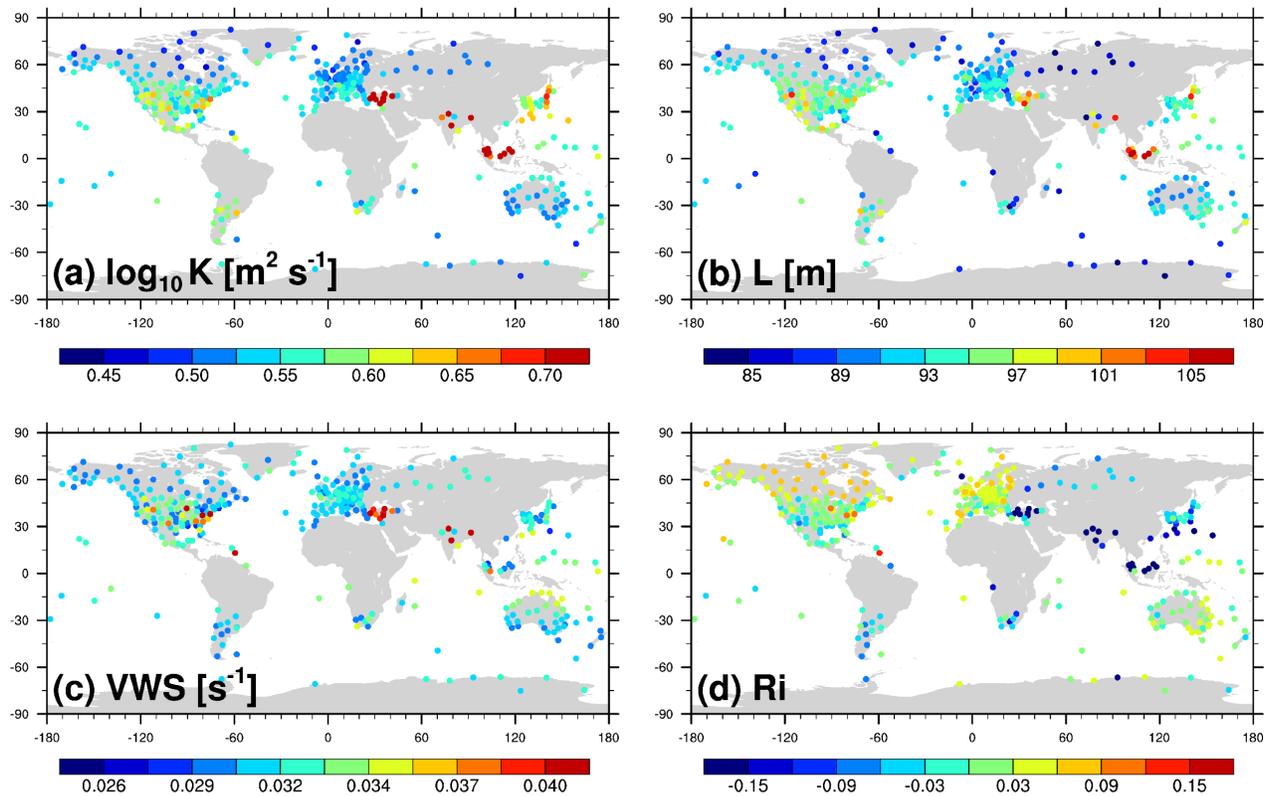

**Fig. 2. Horizontal distributions of stratospheric turbulent diffusivity and related variables.** **(A)** $\log_{10}K$, **(B)** $L$, **(C)** VWS, and **(D)** Ri, averaged from the tropopause to 30 km during October 2014–December 2025.

Atmospheric turbulence can be generated by a variety of sources. Typical sources include strong shear instability near jet streams (26, 27), breaking of mountain-induced gravity waves (28), and convection-related processes such as overturning and the breaking of gravity waves mechanically forced near convective cloud tops (29, 30, 31). Although direct convective overturning is rare in the statically stable stratosphere, gravity waves generated near convective cloud tops can propagate vertically over substantial distances and subsequently break in the stratosphere, thereby producing turbulence (30, 31).

As proxies for mountain-wave turbulence, Fig. 3 shows the horizontal distributions of station height (Fig. 3a) and the zero-wind ($U_0$) ratio (Fig. 3b). Mountain waves are generated by airflow displaced over topography and can propagate upward into the stratosphere. When such waves encounter a background flow with a phase speed close to that of the source topography, that is, near a $U_0$ condition, they can become unstable and break, producing mountain-wave turbulence. In this study, a $U_0$ condition is regarded as satisfying at least one of the following criteria is met within the observed turbulent layer: (1) reversal of zonal wind ($u$) direction, (2) reversal of meridional wind ($v$) direction, or (3) horizontal wind speed ($\sqrt{u^2 + v^2}$) smaller than 0.1 m s$^{-1}$. The fraction of all turbulence cases satisfying at least one of these conditions is defined as the $U_0$ ratio.

Fig. 3(a) shows that stations at relatively high elevations are located over major mountainous and elevated regions, including the Rocky Mountains in the United States, Turkey, and southern Africa. In Fig. 3(b), the $U_0$ ratio is generally larger at low latitudes and smaller at middle and high latitudes. This latitudinal contrast is likely related to the reversal of background winds associated with the quasi-biennial oscillation (QBO) in the tropics and subtropics, whereas westerlies are more persistent at middle and high latitudes. This interpretation is also consistent with Atlas *et al.* (32), who showed that the QBO produces a pronounced zero-wind region in the equatorial stratosphere, within which turbulence occurs frequently. Turkey exhibits a particularly pronounced maximum in the $U_0$ ratio, suggesting that, together with its elevated terrain, it is a favorable region for the generation and breaking of mountain gravity waves. The Rocky Mountains also show a relatively enhanced $U_0$ ratio compared with surrounding regions, indicating a similar interpretation. In India, station elevations are not especially high, but the $U_0$ ratio is relatively large. Considering the substantial horizontal drift of radiosondes (~79 km; 21), this may indicate that the observed turbulence was influenced by gravity-wave breaking above the nearby Himalayan region. In addition, the enhanced turbulence over India may also be related to strong upwelling and deep convection associated with the summer monsoon circulation, which is known to modulate UTLS dynamics and lower-stratospheric gravity-wave activity and turbulence (33). By contrast, southern Africa shows high station elevations but comparatively low $U_0$ ratios, implying that mountain-wave breaking may not be the dominant turbulence source there.

To examine convective influence, convective proxies were derived through the following procedure. First, the presence of a moist-saturated layer in the troposphere was identified using the method of Zhang *et al.* (34), which diagnoses such layers based on whether relative humidity exceeds empirical altitude-dependent thresholds. When a moist-saturated layer was present in the troposphere, the temperature at the top of that layer was used to define a convective profile (CP): A profile was classified as CP$_{210K}$ when the layer-top temperature was lower than 210 K, and as CP$_{235K}$ when it was lower than 235 K. The 210 K threshold has been widely used as a proxy for deep and intense convection, often associated with overshooting convection near the tropopause (35, 36), whereas the 235 K threshold has often been used as a proxy for cold high cloud (36) and thick cirrus shield in the upper troposphere (35). Finally, the ratios of $CP_{210K}$ and $CP_{235K}$ profiles to the total number of HVRRD profiles at each station are shown in Fig. 3(c) and 3(d), respectively.

Fig. 3(c) shows that the $CP_{210K}$ ratio is clearly enhanced over low-latitude regions, where deep convection is climatologically active, and is relatively small at middle and high latitudes. Notably, Malaysia, which was identified as one of the major $\log_{10}K$ maxima in Fig. 2(a), exhibits a particularly large $CP_{210K}$ ratio, suggesting that convective activity is likely a major source of turbulence there. In Fig. 3(d), the $CP_{235K}$ ratio is also large in low latitudes, but elevated values additionally appear in high-latitude regions, reflecting the influence of cold high cloud and cloud-shield-related activity. Its relevance to stratospheric turbulence, however, is less likely direct than that of $CP_{210K}$. In short summary, these results suggest that among the major $\log_{10}K$ maxima identified in Fig. 2(a), turbulence over Turkey, India, and the Rocky Mountains is likely associated mainly with mountain-wave breaking, whereas turbulence over Malaysia is more strongly linked to convective activity. More broadly, deep convection appears to be an important turbulence source in low latitudes, while convective activity related to cold high clouds and cloud shields may contribute more substantially at high latitudes.

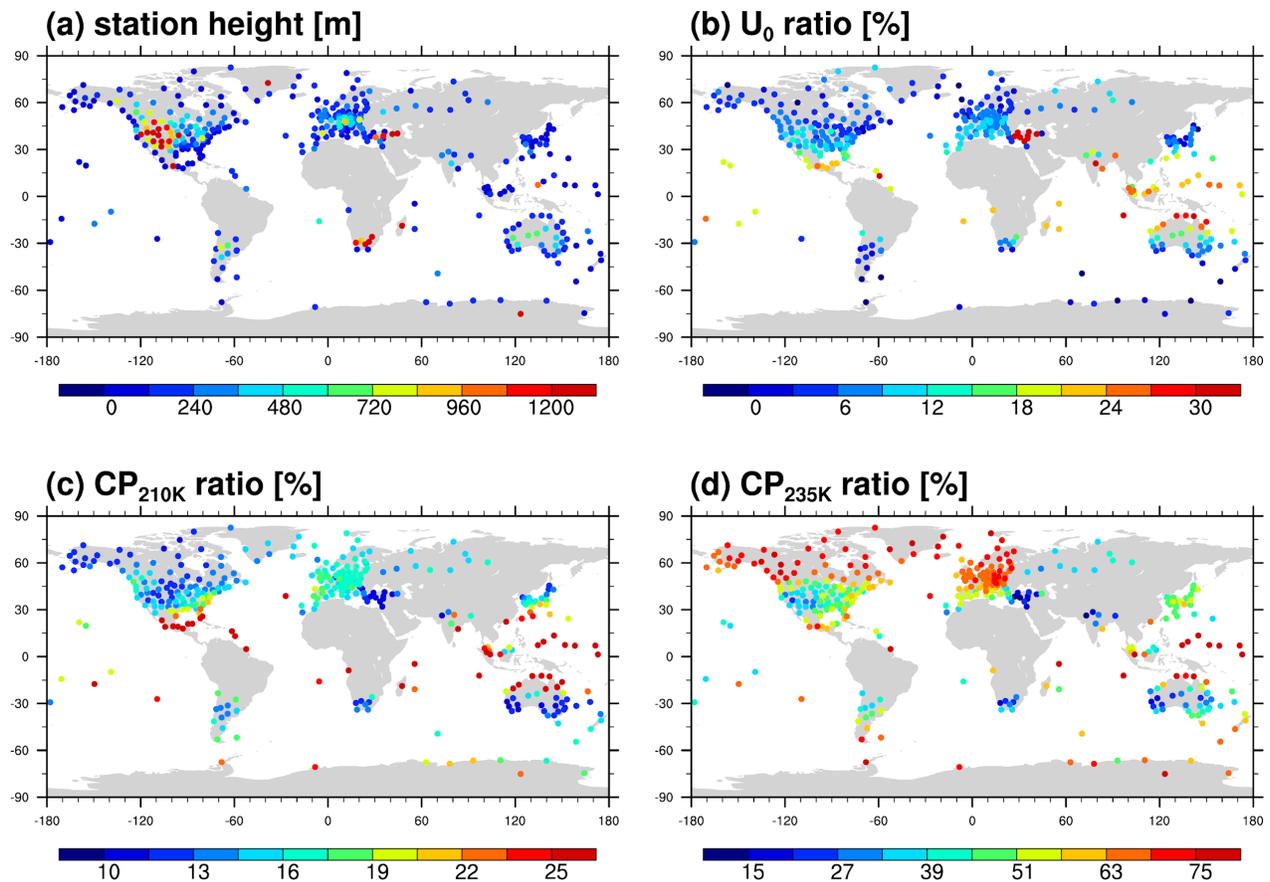

**Fig. 3. Horizontal distributions of topographic and convective proxies for stratospheric turbulence. (A)** Station height, **(B)** zero-wind ($U_0$) ratio, **(C)** convective profile (CP) ratio for a

210 K threshold, and **(D)** CP ratio for a 235 K threshold at each station for z = tropopause–30 km during October 2014–December 2025.

Fig. 4 shows the seasonal variation in the horizontal distribution of log10$K$. Overall, the values in low latitudes remain relatively similar across seasons, whereas those in the middle and high latitudes exhibit a seasonal dependence with larger values generally occurring in the winter hemisphere especially in US and Europe. Additional seasonal distributions of $L$, VWS, and Ri (Fig. S1 in the Supplementary Materials) show that this seasonal variation of log10$K$ corresponds more closely to that of $L$ than to that of VWS. Since $K$ is proportional to the square of $L$ (see Eq. 1 in Materials and Methods), the seasonal modulation of $L$ can have a larger effect on log10$K$ than the corresponding modulation of VWS.

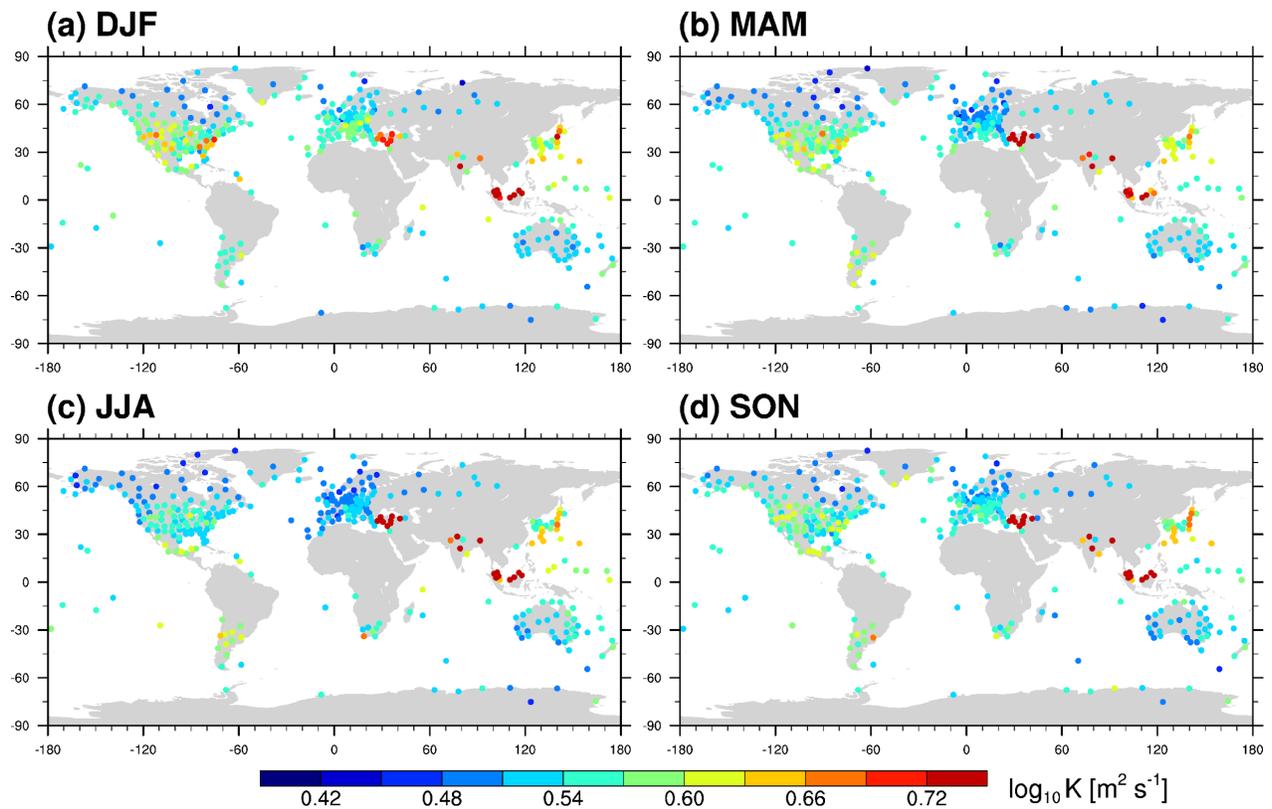

**Fig. 4. Seasonal horizontal distributions of stratospheric turbulent diffusivity. (A)** December–February (DJF), **(B)** March–May (MAM), **(C)** June–August (JJA), and **(D)** September–November (SON) distributions of $\log_{10}K$ at each station for z = tropopause–30 km during October 2014–December 2025.

**2.3. Latitude–altitude structure of stratospheric turbulence**

Fig. 5 presents the latitude–altitude distributions of zonal-mean log10$K$, $L$, VWS, and Ri. For this analysis, 15° latitude bins and 1 km vertical bins were used. It should be noted that caution is required when interpreting these zonal-mean structures, because the inhomogeneous distribution of radiosonde stations inevitably causes some longitude sectors to contribute more strongly than others. Nevertheless, the zonal-mean zonal wind and temperature structures derived from the same radiosonde dataset are highly consistent with well-established large-scale climatological features (Fig. S2 in the Supplementary Materials). Therefore, despite the longitudinal inhomogeneity of the observing network, presenting the zonal-mean stratospheric turbulence structure based on all available observations remains meaningful and informative, especially in the stratosphere where zonal asymmetry is relatively weak.

In Fig. 5(a), $\log_{10}K$ exhibits a pronounced maximum above z = 27 km (within the ozone layer) over nearly all latitudes, with especially large values at 45°S and the Equator–45°N. Below this level, $\log_{10}K$ is relatively uniform with height within the stratosphere, but decreases sharply across the tropopause into the troposphere. Another notable feature is a local maximum just above the tropopause near the Equator–15°N (coinciding with the climatologically strongest ITCZ), centered around z = 17 km. These features can be interpreted more clearly by examining the zonal-mean distributions of the variables contributing to $K$.

The distribution of $L$ in Fig. 5(b) shows a clear contrast between the troposphere and the stratosphere, with relatively large values in the troposphere and smaller values in the stratosphere. This behavior is consistent with the stronger static stability of the stratosphere, and supports the inverse relationship between turbulence-layer thickness and static stability. Within the stratosphere, a vertically elongated local maximum appears near the Equator. This feature coincides with the latitude band of the highest tropopause and may reflect the upward influence of tropical convection, possibly through convectively generated gravity waves in the troposphere and their subsequent breaking in the stratosphere.

The zonal-mean VWS field in Fig. 5(c) shows relatively weak values in the troposphere, intermediate values in most of the stratosphere, and a distinct enhancement above z = 27 km. This indicates that the large $\log_{10}K$ values in the upper stratospheric portion of Fig. 5(a) are closely associated with strong VWS. A plausible explanation is the dynamical influence of the polar vortex and ozone-related thermal structure. Previous studies have suggested that ozone depletion and the associated polar-stratospheric cooling strengthen the polar vortex through thermal-wind balance, thereby enhancing the vertical shear of the zonal wind in the lower and middle stratosphere (37, 38). In the tropical stratosphere, ozone has also been linked to VWS variability through radiative–dynamical coupling associated with ozone transport and the QBO. Dunkerton (39) proposed that vertical advection of ozone in the equatorial lower stratosphere can induce local heating, thereby influencing the circulation and wind structure, and Naoe *et al.* (40) further highlighted the importance of ozone–QBO coupling for the vertical shear structure in the equatorial stratosphere. Although these studies do not directly document the enhanced VWS identified in the present HVRRD analysis above z = 27 km, they suggest that the observed extratropical and tropical enhancements may be qualitatively consistent with the effects of polar-vortex dynamics and ozone-related radiative–dynamical coupling, respectively.

Finally, Fig. 5(d) shows that Ri is generally smaller in the troposphere and larger in the stratosphere. Above z = 27 km, however, Ri decreases again on average, which is likely attributable to the strong VWS seen in Fig. 5(c). Outside this region, the spatial structure of Ri is closely related to those of $L$ and $K$, suggesting that below z = 27 km the distributions of $L$ and Ri play dominant roles in determining the structure of $\log_{10}K$, as already implied by Fig. 2. In contrast, above z = 27 km, the enhanced VWS appears to be the primary factor controlling the zonal-mean distribution of $\log_{10}K$.

The local maximum of $\log_{10}K$ at the Equator–15°N and z = 17 km in Fig. 5(a) may have an important implication for stratospheric aerosol injection (SAI). Because the efficiency of SAI depends strongly on the choice of injection location (4, 5, 6), identifying a dynamically favorable region is of practical interest. From the perspective of enabling faster aerosol dispersion, the present results suggest that the Equator–15°N region near z = 17 km may be a favorable candidate if a single injection location is to be selected. Although the largest $\log_{10}K$ values occur above z = 27 km, such altitudes are not readily accessible to most transport aircraft. In addition, Fig. 5(a) shows another notable feature: a local enhancement of $\log_{10}K$ appears roughly 2 km above the tropopause across a broad range of latitudes. This is consistent with Kim *et al.* (41), who, based on 31 NASA ER-2 flights during the DCOTSS field campaign in the summers of 2021 and 2022, reported that strong turbulence occurs frequently within 2 km above the tropopause, corresponding to about 1–2 km above deep convective cloud tops. This suggests that, if SAI operations were tailored by latitude band, injections performed near this altitude could favor more efficient initial dispersion of injected aerosols. Such enhanced turbulent mixing may help accelerate the early-stage spreading of aerosol plumes after release (7). It should be emphasized, however, that the effectiveness of SAI is not determined by turbulent diffusion alone. Mean-wind advection, radiative effects, microphysical evolution, and removal processes such as deposition must also be considered (5, 7, 42, 43). Therefore, the optimal injection location should ultimately be determined on the basis of these factors together with the turbulent-diffusion characteristics identified in the present study.

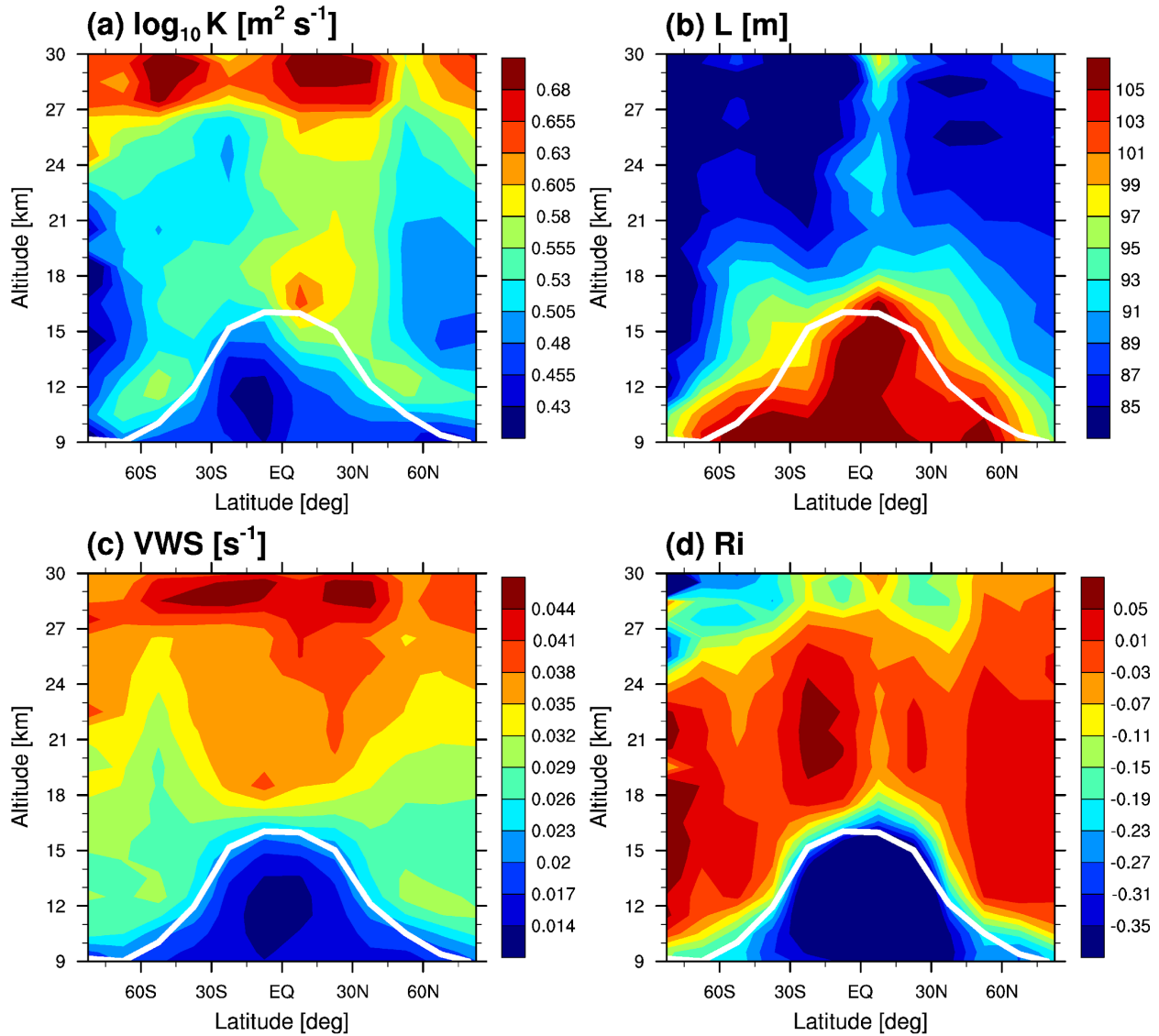

**Fig. 5. Latitude–altitude distributions of zonal-mean stratospheric turbulence characteristics.** **(A)** $\log_{10}K$, **(B)** $L$, **(C)** VWS, and **(D)** Ri during October 2014–December 2025. White lines indicate the tropopause height.

Fig. 6 shows the seasonal variation in the latitude–altitude distribution of zonal-mean log10$K$. The most prominent maximum occurs above $z = 27$ km and is especially strong in the winter hemisphere. This seasonal asymmetry is likely related to the seasonal evolution of the stratospheric polar vortex, which forms and strengthens during winter and weakens during summer, thereby enhancing the background westerlies and associated VWS in the winter stratosphere (44). Another robust feature is that, as in Fig. 5(a), a local maximum of log10$K$ appears approximately 1–2 km above the tropopause, following the seasonal variation of tropopause height. For example, the near-tropopause maximum over the Equator–15°N appears near 17 km in DJF, but shifts to around 16 km in JJA, consistent with the seasonal change in tropopause altitude shown by the white line. Thus, while the exact altitude of the local peak varies slightly with season, its occurrence just above the tropical tropopause is maintained throughout the year. This result also supports the SAI-related discussion based on Fig. 5(a). In that earlier analysis, the Equator–15°N region near z = 17 km was suggested as a practically favorable injection altitude from the perspective of turbulent diffusion. Fig. 6 indicates that, although the magnitude of $\log_{10} K$ exhibits some seasonal variation, the location of the near-tropopause local maximum remains broadly unchanged. Therefore, the proposed implication for SAI remains valid from a seasonal perspective as well, even though the exact optimal altitude may shift slightly with the seasonal displacement of the tropopause.

This latitude–altitude distribution of $K$ may also be viewed in the context of a previous effective-diffusivity study of large-scale atmospheric transport (9). In the lower stratosphere, Haynes and Shuckburgh (9) showed that local minima in effective diffusivity identify tropopause-related transport barriers and that these barriers are strongest in winter and weakened by monsoon circulations in summer. Although their effective diffusivity represents large-scale isentropic tracer mixing rather than the local turbulent diffusivity estimated here, both studies highlight the pronounced seasonal cycle and spatial inhomogeneity in atmospheric mixing.

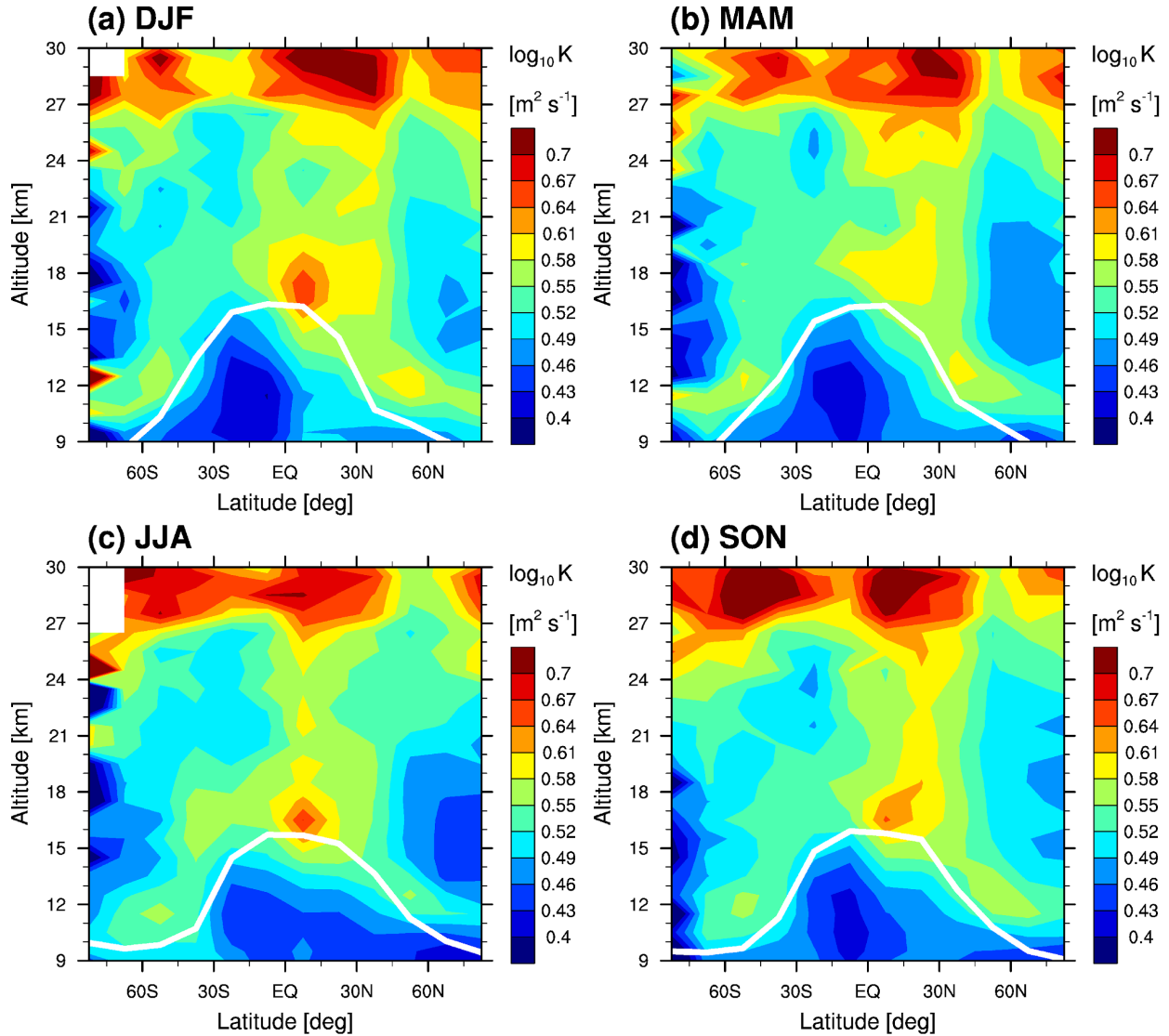

**Fig. 6. Seasonal latitude–altitude distributions of zonal-mean stratospheric turbulent diffusivity. (A)** December–February (DJF), **(B)** March–May (MAM), **(C)** June–August (JJA), and **(D)** September–November (SON) distributions of $\log_{10}K$ during October 2014–December 2025. White lines indicate the tropopause height.

## 2.4. Trends in stratospheric turbulence over the past decade

Fig. 7 presents the temporal evolution of the yearly mean stratospheric turbulence characteristics during 2015–2025. Although this 11-year period is short to establish robust long-term climate trends, it is nevertheless worthwhile to examine how the globally averaged turbulent diffusivity and its governing variables have changed over the available analysis period. As shown in Fig. 7(a), $\log_{10}K$ exhibits a clear increasing trend over the study period. A similar increase is found in $L$ (Fig. 7b), whereas VWS (Fig. 7c) and Ri (Fig. 7d) show comparatively weak temporal changes. For log10$K$ and $L$, statistically significant positive trends are found for all cases, PosRi cases, and NegRi cases at the 95% confidence level. A comparison of the year-to-year variations in log10$K$, $L$, and VWS suggests that the long-term increase in log10$K$ is primarily associated with the increase in $L$, while interannual fluctuations in VWS appear to modulate the detailed temporal evolution of log10$K$, which is also supported by their positive correlations. The correlation coefficient between log10$K$ and VWS is 0.70 for all cases, 0.44 for PosRi cases, and 0.57 for NegRi cases; the correlations are statistically significant for all cases and for NegRi cases at the 95% confidence level. In other words, the increase in log10$K$ is largely controlled by the thickening of turbulence layers, whereas VWS plays a secondary role in shaping shorter-term variability.

Another notable feature is the change in Ri regime composition shown in Fig. 7(e): the fraction of PosRi cases increases significantly over time, whereas that of NegRi cases decreases correspondingly. At the same time, despite the weak increasing tendency in VWS, Ri also shows a slight increase. This implies that static stability must also have increased on average, suggesting a gradual strengthening of static stability in the stratosphere. In this sense, the increasing fraction of PosRi cases may reflect an increasingly stable stratosphere, within which turbulence more frequently occurs under statically stable but shear-unstable conditions. Such an interpretation is physically consistent with a recent study showing continued cooling of the lower stratosphere under climate change (45).

A key question, then, is why $L$ and $\log_{10}K$ also increase despite the apparent increase in static stability. A plausible interpretation is that the turbulence analyzed in this study is restricted to burst-like, active turbulence satisfying Ri < 0.25. Under a more strongly stabilized atmosphere, only turbulence events strong enough to overcome that enhanced stability are likely to be detected within this regime. Accordingly, the positive trends in $L$ and $\log_{10}K$ may indicate that the turbulence events are becoming vertically deeper and more intense. In other words, climate-change-related stratospheric cooling and stabilization may favor a larger proportion of PosRi turbulence, and the turbulence that develops within this increasingly stable environment tends to be stronger. This interpretation is also consistent with the fact that both PosRi and NegRi cases individually exhibit positive trends in $L$ and $\log_{10}K$, indicating that the increase in all-case diffusivity is not solely a result of the increasing PosRi fraction, but also reflects within-regime intensification of turbulence structure.

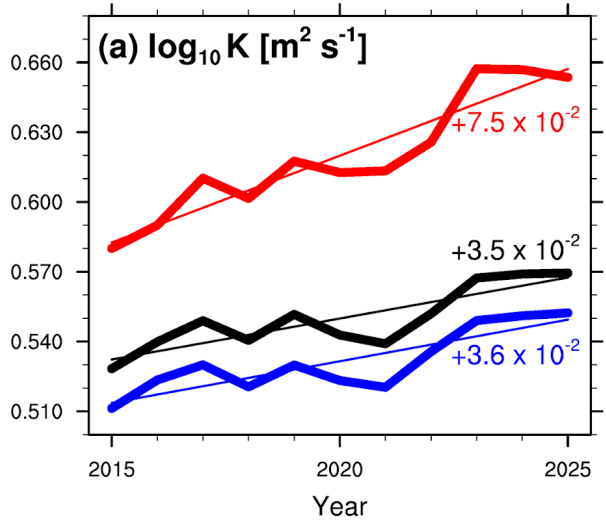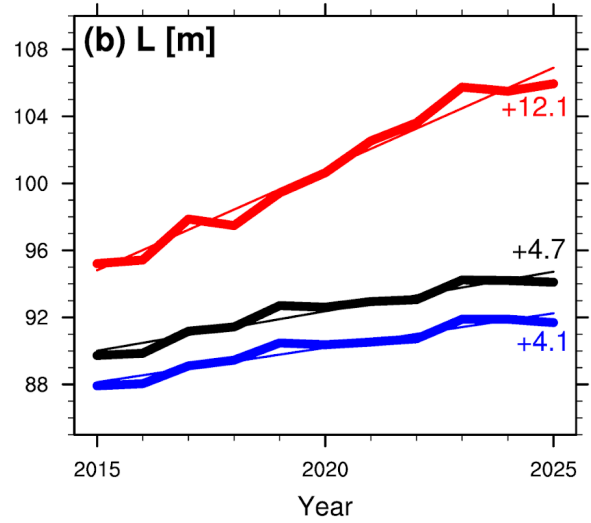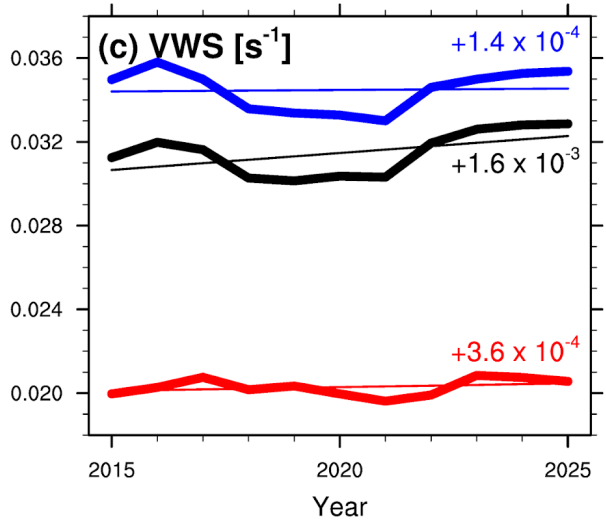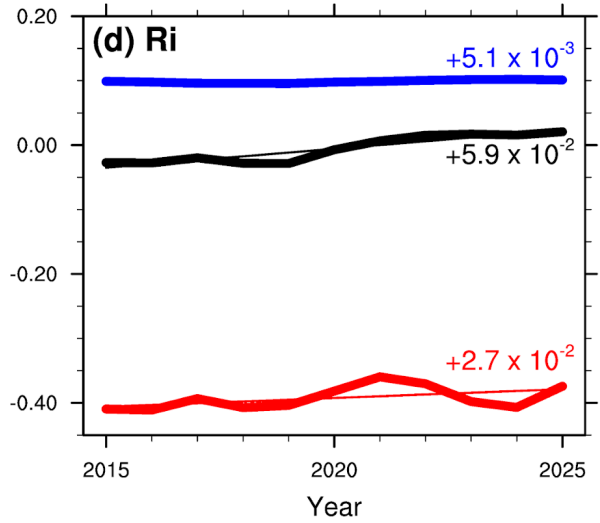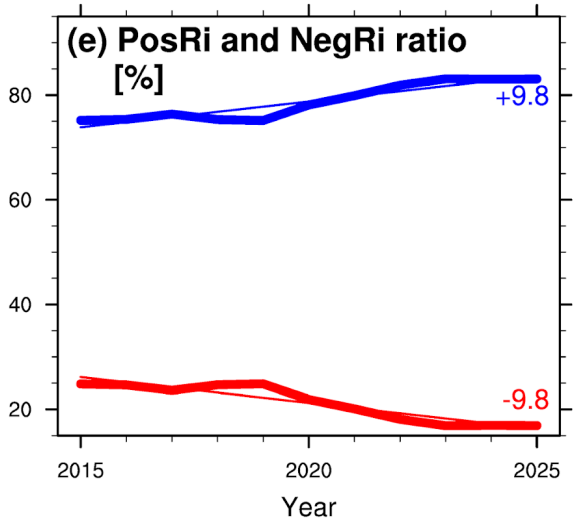

**Fig. 7. Temporal evolution of stratospheric turbulence characteristics. (A)** Yearly mean $\log_{10}K$, **(B)** $L$, **(C)** VWS, **(D)** Ri, and **(E)** ratios of positive-Ri and negative-Ri cases for z = tropopause–30 km during 2015–2025. Thick black, blue, and red lines represent all cases, positive-Ri cases, and negative-Ri cases, respectively; thin lines indicate the corresponding linear trends. Numbers denote the changes over a decade derived from the linear trends.

## 3. Discussion

This study investigated the spatial and temporal variation of turbulent diffusivity in the stratosphere using global operational HVRRD and the $Ri_{min}$-based turbulence estimation method (see detailed description in the Section of Materials and Methods). The analysis used 1,614,270 radiosonde profiles from 370 stations worldwide during October 2014–December 2025 and focused on stratospheric turbulence between the tropopause and 30 km altitude. By applying the $Ri_{min}$-based framework, the present analysis was able to represent turbulence not only in statically unstable conditions but also in statically stable layers with sufficiently strong VWS, which is particularly important in the stratosphere.

The main findings can be summarized as follows. First, the global occurrence distribution of stratospheric turbulent diffusivity spans a broad range, with mean and median $\log_{10}K$ values of 0.554 and 0.490, respectively. Most stratospheric turbulence cases belong to the positive-Ri regime, indicating that shear-driven turbulence in statically stable layers dominates over overturning turbulence. However, although negative-Ri cases are much less frequent, they tend to exhibit stronger diffusivity and larger layer thickness. Second, the horizontal distribution of $K$ shows pronounced regional structure, with maxima over Turkey, India, Malaysia, Japan, and several mountainous regions. Comparisons with $L$, VWS, and source proxies suggest that mountain-wave breaking is an important contributor over Turkey, India, and the Rocky Mountains, whereas convective activity likely plays a larger role over Malaysia and broadly across low latitudes. Third, the seasonal distribution of $K$ is relatively weak in the tropics but shows clear winter enhancement in the extratropics, consistent with stronger seasonal wind shear in the winter hemisphere.

Fourth, the latitude–altitude structures reveal two robust features: a pronounced enhancement of $K$ above z = 27 km and a secondary local maximum just above the tropopause near the Equator–15°N around 17 km altitude. The former appears to be closely associated with enhanced VWS, while the latter is likely related to tropical convective influence and gravity-wave activity. This near-tropopause local maximum is of potential practical interest because, from the perspective of turbulent diffusion, it may represent a favorable SAI injection region for promoting faster initial aerosol spreading. In addition, the presence of a local enhancement approximately 1–2 km above the tropopause across a broad range of latitudes suggests that latitude-dependent injection altitudes may be advantageous if efficient initial aerosol diffusion is desired for SAI. Such implications should, however, be interpreted cautiously, because the effectiveness of SAI is not controlled by turbulent diffusion alone but also by advection, radiative processes, aerosol microphysics, and removal processes.

Finally, the annual means for 2015–2025 show significant positive trends in both $K$ and $L$, together with a significant increase in the fraction of positive-Ri cases. Although the available record is still too short to establish robust long-term climate trends, these results suggest that stratospheric turbulence may be evolving over the recent observational period. The increasing fraction of positive-Ri cases is consistent with a gradually more stable stratospheric background, while the simultaneous increase in $L$ and $K$ implies that turbulence events occurring within this environment may be becoming vertically deeper and more intense. In this sense, the recent evolution of stratospheric turbulence may involve both a shift in regime composition and a within-regime intensification of turbulence structure.

Overall, this study provides observationally based reference information on stratospheric turbulent diffusivity from a near-global radiosonde network. The results help constrain the magnitude, spatial variability, and temporal behavior of stratospheric turbulent diffusion, and thus offer useful guidance for turbulence parameterization (especially for plume and box models), stratospheric transport and mixing studies, and future discussion of aerosol dispersion in SAI-related applications. At the same time, several limitations should be acknowledged. Because HVRRD reflect the atmosphere after turbulence generation has already occurred, they are less suitable for diagnosing the full background dynamical conditions responsible for turbulence initiation. In addition, the zonal-mean and global structures derived here are affected by the spatial inhomogeneity of radiosonde stations, and the recent temporal trends should be interpreted with caution given the still limited record length. Future work combining HVRRD with reanalysis or model datasets would help further clarify the large-scale environments controlling the observed turbulence distributions and trends.

## 4. Materials and Methods

### 4.1. Data: High vertical-resolution radiosonde data (HVRRD)

This study uses global operational HVRRD provided by the European Centre for Medium-Range Weather Forecasts (ECMWF) through the U.S. National Centers for Environmental Information (NCEI). These observations are collected within the operational radiosonde network and exchanged internationally through the World Meteorological Organization (WMO) Information System (18). The dataset includes in situ measurements of fundamental atmospheric variables, such as temperature, pressure, humidity, and horizontal wind, together with position information along the balloon trajectory. Only radiosonde data reported at 1 and 2 s intervals were used in this study, corresponding to approximate vertical resolutions of 5 and 10 m, respectively. Such fine vertical sampling makes the dataset suitable for investigating small-scale structures in the atmosphere.

In the present study, the analysis covers 11 years and 3 months from October 2014 to December 2025, and includes data from 370 radiosonde stations worldwide with launches generally available at 00 and 12 UTC. It should be noted that the data availability varies among stations (see Figure S1 of 16). A total of 1,614,270 radiosonde profiles from 370 stations were analyzed during the analysis period. As with previous studies using this dataset (18, 19), the spatial and temporal sampling reflects the heterogeneous nature of the operational global radiosonde network, denser coverage over continental midlatitudes, and sparser sampling over oceans and some tropical and polar regions.

Because radiosonde ascent rates vary, the vertical spacing is not uniform even for a fixed reporting interval. To obtain consistent vertical grids, the 1 and 2 s data were interpolated to uniform intervals of 5 and 10 m, respectively. A further practical issue is contamination by instrumental noise and balloon pendulum motion, both of which can introduce artificial small-scale fluctuations in temperature and wind profiles. Following the procedure adopted in Ko and

Chun (19), a moving-average filter corresponding to a 60 m vertical scale was applied after interpolation in order to reduce these effects. Specifically, 13-point and 7-point moving averages were used for the 1 and 2 s data, respectively, and the smoothing was applied in a centered manner. This treatment improves consistency between datasets with different reporting intervals, although turbulence layers thinner than 60 m cannot be resolved. A more detailed description of the dataset and processing procedures is provided in Ko and Chun (19).

**Turbulence estimation: Minimum Richardson number ($Ri_{min}$) method**

To estimate turbulent diffusivity from HVRRD, this study employed the $Ri_{min}$-based framework of Ko and Chun (19). $Ri_{min}$ represents the lowest attainable Richardson number when wave-induced perturbations to static stability and vertical wind shear are taken into account (28). This approach can represent turbulence both in statically unstable conditions (Ri < 0) and in statically stable layers with strong wind shear (0 < Ri < 0.25), where Kelvin–Helmholtz instability may develop. Following Ko and Chun (19), the Richardson number calculated directly from HVRRD is used as the $Ri_{min}$. This is based on the assumption that the observed Ri already incorporates the effects of wave-induced perturbations resolved in the radiosonde measurements with 5 and 10 m resolutions.

The diffusion coefficient $K$ is estimated using the first-order Smagorinsky closure (1, 19, 31, 46, 47) as

$$K = (0.2L)^2 |Def| \sqrt{0.25 - Ri_{min}} \quad (1)$$

where $L$ is the turbulence length scale, $Def$ is the total deformation.

A practical issue in applying Eq. (1) to HVRRD is the specification of $L$ and $Def$. Following Ko and Chun (19), a turbulence layer is defined as a vertically contiguous layer in which Ri remains smaller than 0.25 at successive levels, and the thickness of this layer is taken as $L$. For each turbulence layer, the layer-representative Ri is defined as the minimum value of Ri within that layer, and this value is subsequently assigned to $Ri_{min}$ in Eq. (1). In addition, turbulence layers are classified into two regimes according to the sign of Ri: layers containing at least one level with Ri < 0 are categorized as negative-Ri (NegRi) cases, whereas the remaining layers are treated as positive-Ri (PosRi) cases. This classification allows turbulence to be interpreted separately for statically unstable conditions and for statically stable but dynamically unstable conditions associated with strong vertical shear.

Another practical consideration is the evaluation of $Def$ in Eq. (1). In principle, total deformation includes both horizontal and vertical shear, but such a full calculation is not feasible with radiosonde observations because horizontal derivatives are not directly available and vertical velocity is not observed. Ko and Chun (19) showed through a scale analysis that, under large- and meso-scale atmospheric conditions that responsible for turbulence generation, the vertical wind shear term is order-of-magnitude larger than the horizontal shear terms, implying that the contribution from horizontal deformation is comparatively small. Therefore, $Def$ is approximated here by the vertical wind shear (VWS) following Ko and Chun (19). This approximation is physically reasonable for large- to meso-scale background flows, although some caution is needed in environments where horizontal deformation may become more important, such as strongly curved or anticyclonic jet streams (19).

Another method widely used to estimate atmospheric turbulence from HVRRD is the Thorpe method (13, 18, 20, 21, 22, 48). This method identifies turbulence from overturning structures in the observed stratification and therefore is inherently limited to cases with static instability. As discussed by Kantha (49) and Ko and Chun (19), however, atmospheric turbulence can also develop in statically stable environments when VWS is sufficiently strong. Because the Thorpe method is based on the detection of overturning, it cannot capture such shear-driven turbulence, especially when overturning has not yet developed. To overcome this limitation, Ko and Chun (19) proposed a turbulence estimation method based on the $Ri_{min}$. Their method can detect turbulence not only in statically unstable conditions but also in statically stable yet dynamically unstable layers. They further showed that turbulence estimated from the $Ri_{min}$ method is spatially and temporally consistent with aircraft turbulence observations, supporting its applicability to HVRRD-based turbulence analysis. For this reason, the present study uses the $Ri_{min}$ method to estimate turbulent diffusivity $K$.

**Acknowledgments**

We thank Rob Wood, Peter Blossey, and Lucas McMichael from the University of Washington for helpful discussions of the results. This work used Bridges-2 at Pittsburgh Supercomputing Center through allocation EES210037 from the Advanced Cyberinfrastructure Coordination Ecosystem: Services & Support (ACCESS) program, which is supported by National Science Foundation grants #2138259, #2138286, #2138307, #2137603, and #2138296.

**Funding**

    Simons Foundation grant SFI-MPS-SRM-00005157 and SFI-MPS-SRM-00012079

    Quadrature Climate Foundation grant 01-21-000430

**Author contributions:**

    Conceptualization: HCK, HS

    Methodology: HCK, HS

    Investigation: HCK, HS

    Visualization: HCK

    Supervision: HS

    Writing—original draft: HCK, HS

    Writing—review & editing: HCK, HS

**Competing interests:** Authors declare that they have no competing interests.

**Data and materials availability**

    The HVRRD used in this study to estimate turbulence are openly available from the National Centers for Environmental Information (https://www.ncei.noaa.gov/data/ecmwf-global-upper-air-bufr/).

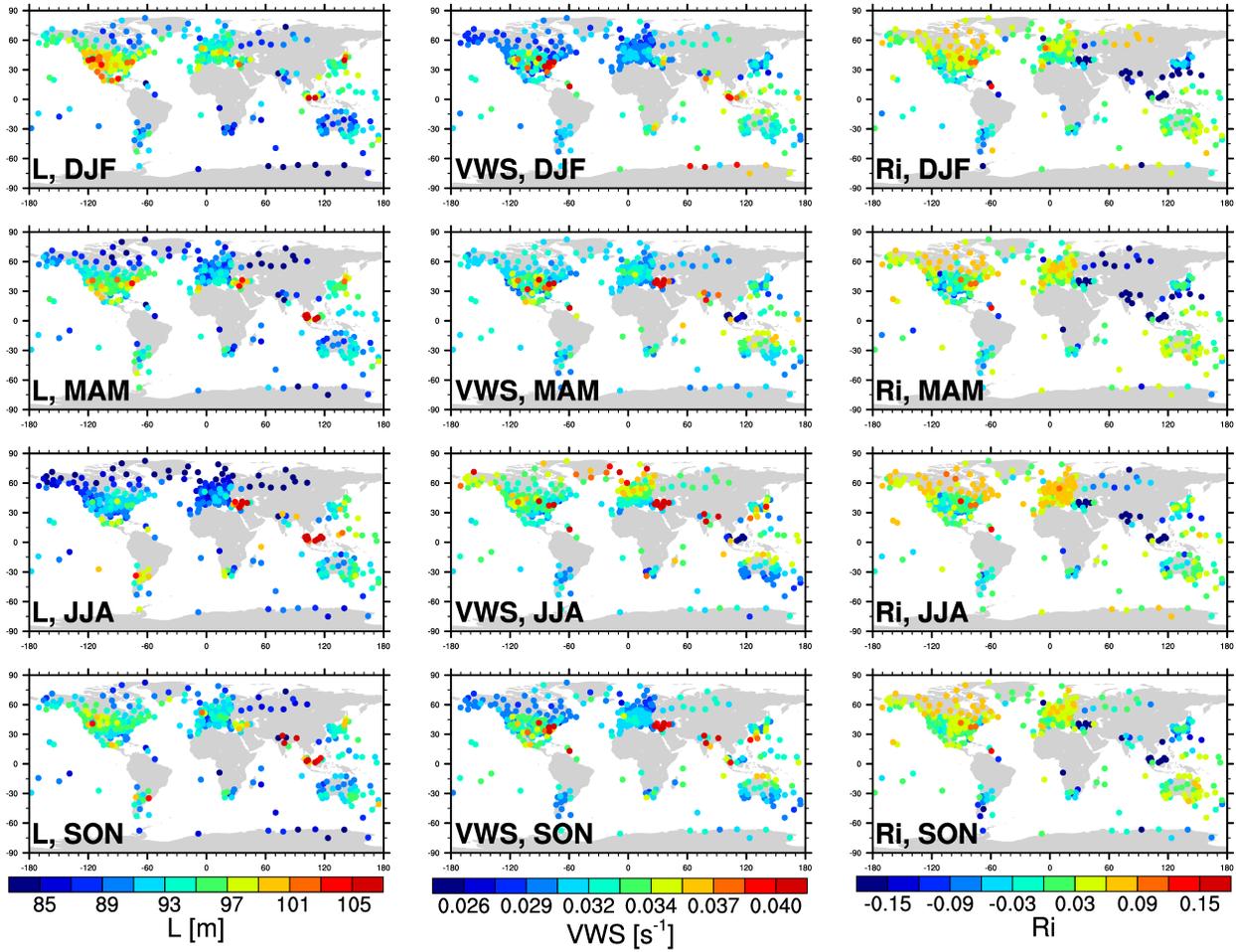

Fig. S1. **Seasonal horizontal distributions of variables related to stratospheric turbulent diffusivity.** From left to right: turbulence length scale (***L***), vertical wind shear (VWS), and Richardson number (Ri); from top to bottom: December–February (DJF), March–May (MAM), June–August (JJA), and September–November (SON). All variables are averaged from the tropopause to 30 km during October 2014–December 2025.

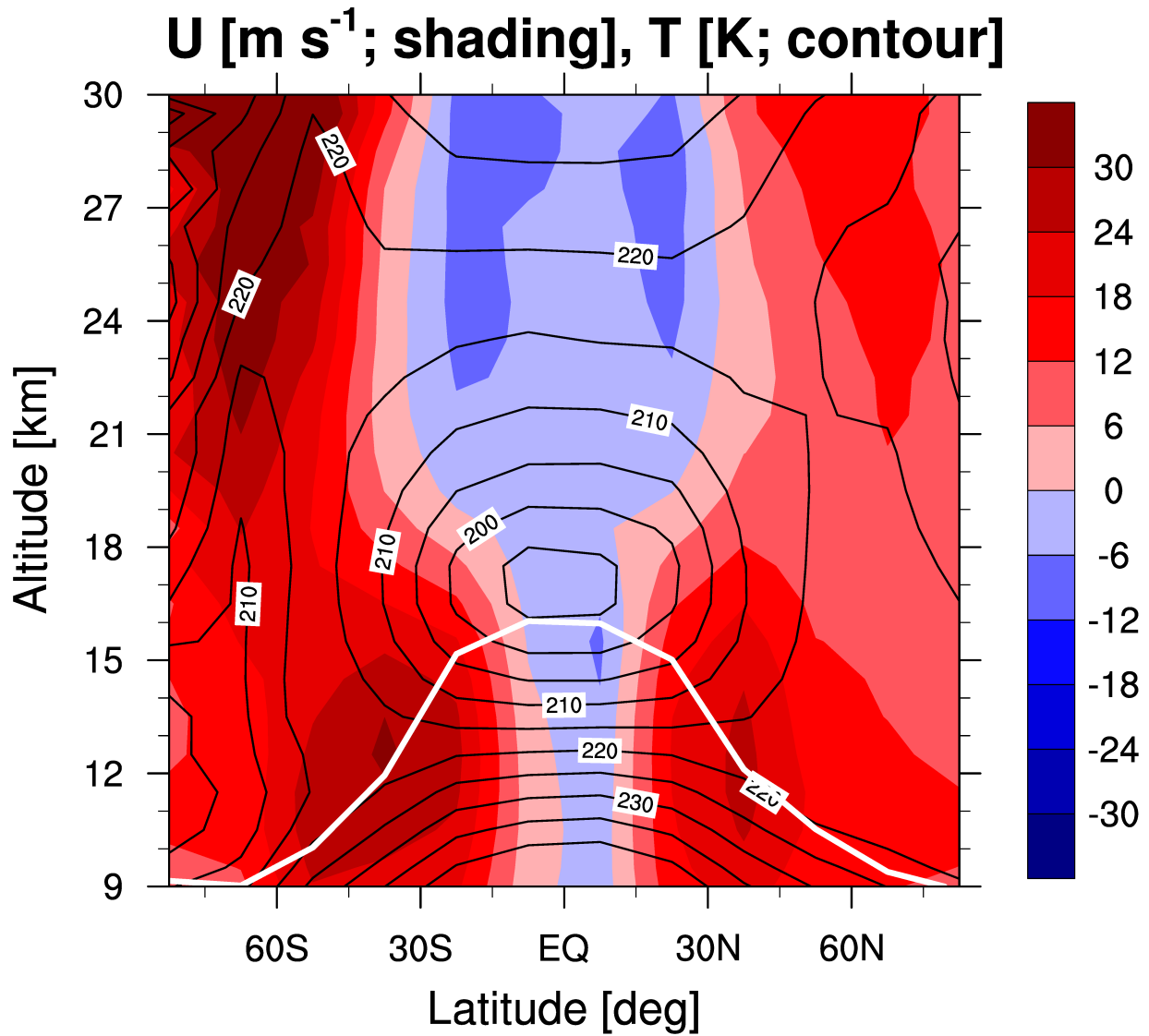

**Fig. S2. Latitude–altitude distributions of zonal-mean zonal wind and temperature observed from high vertical-resolution radiosonde data.** Zonal wind (U) is shown by shading, and temperature (T) is shown by contours during October 2014–December 2025. White lines indicate the tropopause height.